\documentclass[sn-mathphys,Numbered]{sn-jnl}% Math and Physical Sciences Reference Style
\usepackage{graphicx}%
\usepackage{multirow}%
\usepackage{amsmath,amssymb,amsfonts}%
\usepackage{amsthm}%
\usepackage{mathrsfs}%
\usepackage[title]{appendix}%
\usepackage{xcolor}%
\usepackage{textcomp}%
\usepackage{manyfoot}%
\usepackage{booktabs}%
\usepackage{algorithm}%
\usepackage{algorithmicx}%
\usepackage{algpseudocode}%
\usepackage{listings}%
\theoremstyle{thmstyleone}%
\theoremstyle{thmstyletwo}%

\theoremstyle{thmstylethree}%

\usepackage{subfigure}
\usepackage{slashed}
\usepackage{dcolumn}
\usepackage{bm}
\usepackage{epsfig}
\usepackage{amsfonts}
\usepackage{amssymb,amscd}
\def\lsim{\raise0.3ex\hbox{$<$\kern-0.75em\raise-1.1ex\hbox{$\sim$}}}
\def\gsim{\raise0.3ex\hbox{$>$\kern-0.75em\raise-1.1ex\hbox{$\sim$}}}

\newcommand{\be}{\begin{equation}}
\newcommand{\ee}{\end{equation}}
\newcommand{\ba}{\begin{eqnarray}}
\newcommand{\ea}{\end{eqnarray}}

\def\beq{\begin{equation}}
\def\eeq{\end{equation}}
\def\beqa{\begin{eqnarray}}
\def\eeqa{\end{eqnarray}}

\def\gappeq{\mathrel{\rlap {\raise.5ex\hbox{$>$}}
{\lower.5ex\hbox{$\sim$}}}}
\def\lappeq{\mathrel{\rlap{\raise.5ex\hbox{$<$}}
{\lower.5ex\hbox{$\sim$}}}}
\def\Toprel#1\over#2{\mathrel{\mathop{#2}\limits^{#1}}}

\begin{document}

\title[Unveiling the properties of the dimuonium at the energies available at the Large Hadron Collider at CERN]{Unveiling the properties of the dimuonium at the energies available at the Large Hadron Collider at CERN}

\author*[1,2]{\fnm{C.A.} \sur{Bertulani}}\email{carlos.bertulani@tamuc.edu}

\author[2]{\fnm{D.} \sur{Bhandari}}\email{dbhandari@leomail.tamuc.edu}

\author[3]{\fnm{F.S.} \sur{Navarra}}\email{navarra@if.usp.br}
\equalcont{These authors contributed equally to this work.}

\affil*[1]{\orgdiv{Institut f\"ur Kernphysik}, \orgname{Technische Universit\"at Darmstadt}, \orgaddress{\street{Planckstrasse 1},  \city{Darmstadt}, \postcode{64289}, \country{Germany}}}

\affil[2]{\orgdiv{Department of Physics and Astronomy,}, \orgname{Texas A\&M University-Commerce}, \orgaddress{\street{2200 Campbell Street}, \city{Commerce}, \postcode{75429}, \state{TX}, \country{U.S.A.}}}

\affil[3]{\orgdiv{Instituto de F\'{\i}sica}, \orgname{Universidade de S\~{a}o Paulo}, \orgaddress{\street{05315-970}, \city{S\~{a}o Paulo}, \postcode{C.P. 66318}, \state{SP}, \country{Brazil}}}

\abstract{We study the production of the dimuonium (also known as true muonium) in  two and three photon fusion processes in nucleus--nucleus  
collisions at the CERN Large Hadron Collider (LHC) energies. A new formalism is introduced for the production process and valuable new information is extracted which will be helpful in proposals of future experiments.  We explore the phase space constraints, the reaction mechanisms, and how the dimounium decay observables might be jeopardized by other physical processes. We show that the energies available at the large hadron collider at CERN might lead to the first identification of the dimounium in a terrestrial laboratory. }




\maketitle

\section{Introduction}\label{sec1}
There should exist 6 leptonic atoms, (1) the positronium (e$^+$e$^-$), (2) muonium ($\mu^+$e$^-$), (3) dimuonium ($\mu^+\mu^-$), (4) tauonium ($\tau^+$e$^-$), (5) tau-muonium ($\tau^+\mu^-$), and (6) the ditauonium ($\tau^+\tau^-$). Only the positronium, the muonium, and the dipositronium, a ($e^+e^-$)($e^+e^-$)  molecule, have been observed \cite{Deut53,Hug60,CM07}. The dimuonium is more compact than the positronium and muonium: $R_{\mu\mu} \sim R_{\mu e}/100 \sim R_{ee} /200$. Due to its large mass it is sensitive to physics beyond the standard model \cite{FJS23}. The decay rate of the dimuonium depends on radiative corrections from the unexplored timelike region of QED. The observation of the dimuonium \cite{HugM71} would be a significant discovery in physics. It would add to new tests of QED because the properties of the dimuonium are similar to those of the positronium  obtained by just changing the electron by the muon  mass \cite{KJIS98}. It is well known that there are muon sector anomalies, because there is a 3$\sigma$ difference between the muon (g-2) standard model prediction and measurements \cite{Aoy20}. The magnetic moment of each muon in the dimuonium includes an anomalous (g-2) part. Therefore, the hyperfine splitting in the  dimuonium may help clarify the discrepancy between the anomalous magnetic moment of the muon and the theoretical predictions. The (g-2) value for the muon can be modified by various Beyond the Standard Model (BSM) phenomena including leptoquarks, lepton compositeness, Supersymmetry (SUSY), Born-Infeld extensions of QED, spacetime non-commutativity in QED, extra spatial dimensions, etc \cite{KKY22}.  Quantum chromodynamics introduces corrections in the dimuonium structure in the form of relatively well-studied hadronic vacuum polarization. Electroweak corrections to the hyperfine splitting of dimuonium can also be calculated. 

The properties of the dimuonium can shed light on the proton-deuteron radius puzzle \cite{Poh10}, besides of clarifying a possible lepton-universality violation in rare B decays such as the branching ratios for  $B^+ \rightarrow K^+ e^+e^-$ and and $B^+\rightarrow K^+ \mu^+\mu^-$.  The LHCb collaboration has measured this branching ratio and found a 2.5$\sigma$ discrepancy with theoretical predictions \cite{Aaj19}. Finally, BSM physics scale with corrections with the scale parameter $\Lambda$ and manifests itself in the form of corrections of power counting ${\cal O}(m_\mu^2/\Lambda^2)$, making the dimuonium more than $10^5$ more sensitive to BSM than the positronium or the muonium.

The production of the dimuonium in $e^+e^-$ and heavy ion collisions has been studied theoretically in, e.g., Refs. \cite{BS62,Mof75,GJK98,BL09,AGM20}. The bound states are the spin-singlet paradimuonium (PM) $^1$S$_0$ ($J^{PC}= 1^{--}$) and the spin-triplet orthodimuonium (OM) $^3$S$_1$ ($J^{PC}= 0^{-+}$). The leading order prediction for the $^3S_1-^1S_0$ hyperfine splitting is $\Delta E = 7m_\mu \alpha^4/12$ \cite{JPG97}.  The PM decays into two gammas with a lifetime $\tau_{PM}=2\hbar n^3/\alpha^5 m_\mu=0.602$ ps while the OM decays into an $e^+e^-$ pair with $\tau_{OM}=6\hbar n/\alpha^5 m_\mu= 1.81$ ps. Therefore, the 2.2 $\mu$s muon weak decay lifetime is effectively infinite when compared to the dimuonium lifetime.  On the other hand, the ditauonium annihilation decay and the weak decay of the tau lepton are of the same order of magnitude and the bound $\tau^+\tau^-$ system is not as suitable as the dimuonium for precision QED tests. Moreover, the small signal yield compared to the background decreases largely the prospects for the ditauonium being observed at LHCb \cite{LYJ18,Vid19,ES23}.
There are recent experimental proposals to produce the dimuonium at the LHCb as a displaced $e^+e^-$ resonance or via rare B decays into leptonium \cite{Fae18,Vid19}, in low energy $\mu^+\mu^-$ collider \cite{Dru18}, or in $e^+e^-$ collisions  at the Fermilab by a modification of the existing FAST/NML facility \cite{FJS23}. The direct production of dimuonium in $\mu^+\mu^-$ collisions meets the experimental difficulties associated with the production of slow muon beams. Most of the proposals exploit the production of the OM ($^3S_1$) and its decay via the detection of the $e^+e^-$ pairs. The production of PM ($^1S_0$) is often neglected because its dominant decay into $\gamma\gamma$ is challenging to reconstruct with present detectors \cite{Fae18,Vid19,FJS23}. 

In this work we focus on the possibility that the properties of the PM can be studied with the energies presently available at the Large Hadron Collider (LHC) at CERN.  We propose a new venue to calculate the PM production probabilities and study its dependence on phase-space components. In heavy ion collisions, such as Pb + Pb, the production cross sections  of the OM are orders of magnitude smaller than the production of the PM because they require the fusion of  three photons as in the case of the production of the  orthopositronium \cite{bn02}.   Our goal in this work is to quantify the regions in phase space where the dimuonium can be observed in ultra-peripheral collisions (UPC) of relativistic heavy ions \cite{BB88a}. We  adopt a projection technique  \cite{bn02} to calculate tree-level Feynman diagrams and connect the calculated amplitude with the  production of a bound PM. This contrasts with the frequently used virtual photon approximation, which relies on a folding of virtual photon fluxes from each ion multiplying a cross section for the gamma-gamma particle production \cite{BB88a}.   
The production of free $\mu^+\mu^-$ and $\tau^+\tau^-$ pairs at the LHC has been predicted and calculated following this recipe in Ref. \cite{BB87}.  Alternatively, the production of exotic atoms consisting of muons and taus in the atomic nucleus orbitals at the LHC was calculated using a direct application of perturbative QED in Ref. \cite{BE10}.    

\section{Dimuonium probability distribution} 
In this work we adopt the laboratory system of the collider and a first-order perturbation theory for the dimuonium production, well justified due to its small creation probability. The direct and exchange Feynman diagram added together yield the  probability amplitude for the production of a $\mu^+\mu^-$ pair as given by
\beqa
{\cal M}&=&-ie^2\bar{u}\left({P\over 2}\right)\int{d^4q \over {(2\pi)^4}}\slashed{A}_1\left( {P\over 2} -q\right) {{\slashed{q} + m_\mu} \over {q^2 - m_\mu^2 }}\nonumber \\
&\times&\slashed{A}_2\left( {P\over 2} +q\right)v\left( {P\over 2}\right) + \left(\slashed{A}_1 \leftrightarrow \slashed{A}_2\right) , \label{Mamp}
\eeqa 
where the electromagnetic field of ion 1 at the center of mass of the collision with impact parameter ${\bf b}$ is given by
\beq
A^0_1=-8\pi^2 Ze \delta(q_0-\beta q_z) {e^{i{\bf q}_T \cdot {{\bf b} / 2}}\over {q^2_T + q^2_z/\gamma^2}}, \ \ \ A_1^z = \beta A_1^0.
\eeq
Only the time-like (0) and third ($z$) components of the electromagnetic fields are needed, assuming that the quantization axis  $z$ is along the beam direction. The index $T$ denotes the direction perpendicular to the beam. Here the Lorentz factor is given by $(1-\beta^2)^{-1/2}$ in terms of the beam velocity $\beta$. To obtain the field $A_2$ of ion 2, one uses the replacements ${\bf b} \rightarrow -{\bf b}$ and $\beta \rightarrow -\beta$. In Eq. \ref{Mamp}, $P$ is the dimuonium momentum, $m_\mu= 105.66$ MeV/c = 0.5267 fm$^{-1}$ is the muon mass, $Z$ is the ion charge and $u$ ($v$) is the spinor associated with the $\mu^-$ ($\mu^+$). 

The production probabilities and cross sections for the dimuonium rely on Eq. \ref{Mamp} and the projection technique introduced in Ref. \cite{bn02} for the creation of the positronium in UPCs. We consider only the production of the PM ($^1S_0$) via two-photon fusion. When calculating production probabilities and cross sections, we need to average over spins and use the projection operators onto bound dimuonium states, obtained by the replacement  \cite{Nov78,KKS,BYG85}
\beq
\bar{u}\big( \cdots \big)v \rightarrow {\Psi(0)\over 2\sqrt{2m_\mu}}{\rm{tr}}\left[ \cdots (\slashed{P} +2m_\mu)i\gamma^5\right], \label{uvproj}
\eeq
where   $\Psi(0)$ is the dimuonium wavefunction at the origin.
This procedure leads to integrals stemming from Eq. \ref{Mamp} that are performed using Feynman's well-known parametrization to solve integrals of products of fractions \cite{Peskin}. Our final expression for the probability ${\cal P}$ of PM production is
\beqa
{d{\cal P}(P_T,P_z,b)\over dP_T dP_z}= 64 m_\mu^3\left| \Psi(0)\right|^2 Z^4 \alpha^4 f^2(C_i;b){P_T \over P_0}, \  \ \ \  \label{dPdm}
\eeqa
where  $P_0 = \sqrt{4m_\mu^2 +P_T^2 + P_z^2}$, and
\beq
f(C_i;b)=\sum_{i=1}^3 a_i K_0(C_ib),  \label{fob}
\eeq
with
\beqa
a_i &=& {1\over {(C_j^2- C_i^2)(C_k^2-C_i^2)}}, \ \ \ {i \ne j\ne k=1,2,3, } \nonumber \\
C_1^2 &=& {P_0^2/\beta^2-\beta^2P_z^2  \over 4}+m_\mu^2 \  ,\ \ C_2^2 = {P_T^2\over 4}+{{P_z^2 + P_0^2/\beta^2}\over {4\gamma^2}} ,\nonumber \\ 
C_3^2 &=& {P_T^2\over 4}+{{P_z^2 - P_0^2/\beta^2}\over {4\gamma^2}}. \label{fob2}
\eeqa
The functions $K_0(C_ib)$ are modified Bessel functions of the second kind and order zero. As shown in Ref. \cite{bn02}, Eq. \ref{dPdm} reduces to the well known equivalent photon approximation (EPA) after additional approximations are used. A comparison of the two methods will be shown later in this article to explore the validity of the equivalent photon approximation. The projection technique implemented via Eq. \ref{uvproj} is a powerful simplification allowing to calculate simple tree-level diagrams and projecting the open legs into bound states.

\begin{figure}[t]
\begin{center}
\includegraphics[
width=3.2in]
{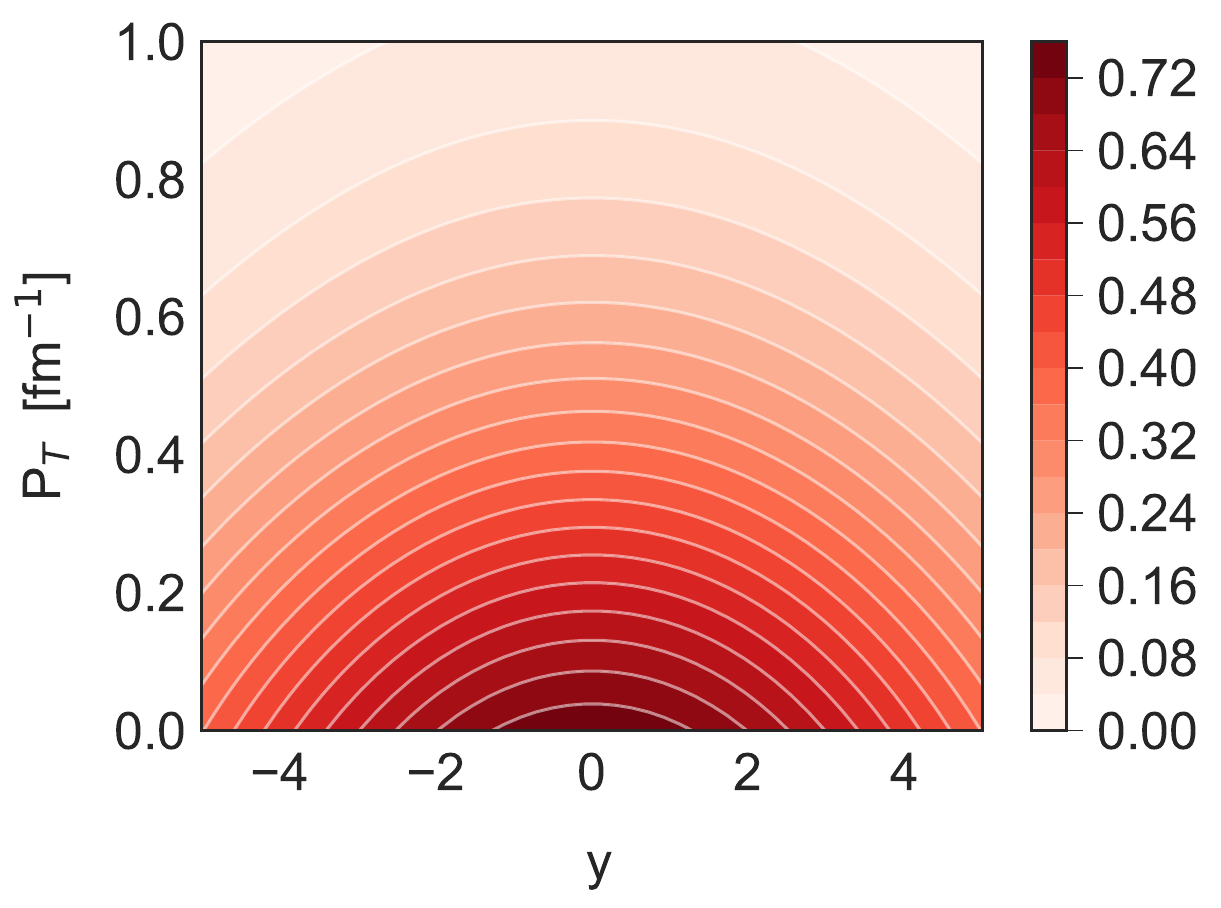}
\caption{Contour plot  of the momentum-rapidity distribution $d{\cal P}_{dm}/ dP_T dy$ (scaled to unity at $P_T =0, \   y=0$) for para-dimuonium production  for a Pb + Pb collisions at the LHC at an impact parameter $b=15$ fm. }
\label{dpdptdy}
\end{center}
\end{figure}

\section{Results}\label{sec2}

The LHC provides  Pb + Pb collisions with center of mass energy of $\sqrt{s_{NN}} = 5.5$ TeV/nucleon, corresponding to a laboratory beam energy of $E_{lab}=2.76$ TeV/nucleon and to a Lorentz factor $\gamma=2941$. We use Eq.  \ref{dPdm} to obtain the distribution probability of dimuonium production events as a function of the impact parameter $b$ and the momentum components $P_z$ ($P_T$) along (perpendicular) the beam direction. The experimental determination of this distribution is a way to access information on the dimuonium wave function $\Psi(0)$.  
Based on a similar treatment as the non-relativistic Schr\"odinger equation, it is given by
$  \Psi(0) = 1/ (na_0)^{3/2}\sqrt{\pi}$ =  $4.8 \times 10^{-5}$ fm$^{-3/2}/n$, where the Bohr-radius for the dimuonium is  $ a_0 = 2 / m_\mu \alpha = 512$  fm (we use $\hbar = c = 1$) and $n$ is the principal quantum number.

\begin{figure}[t]
\begin{center}
\includegraphics[
width=3.in]
{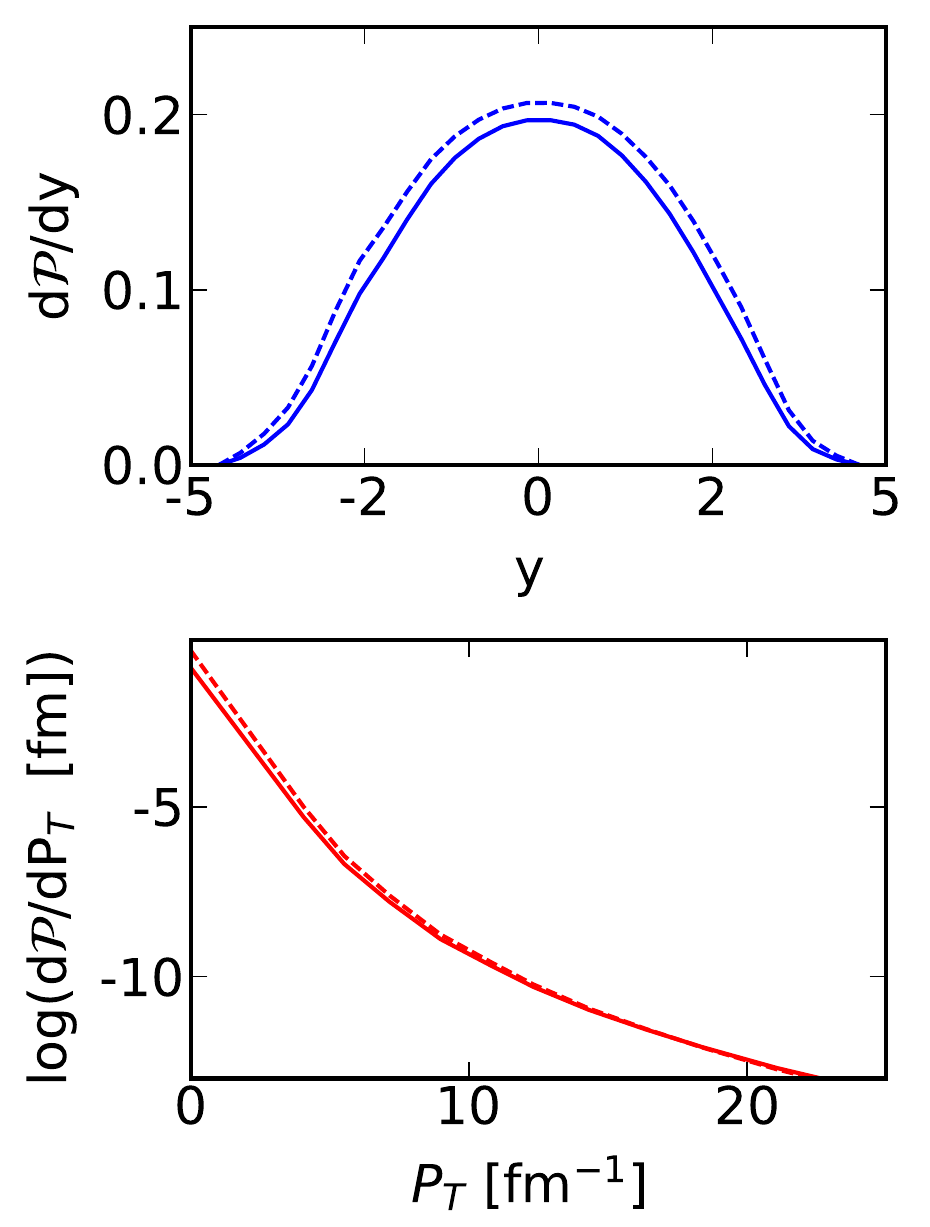}
\caption{{\it Upper Panel:} Rapidity distribution of para-dimuonium produced in Pb + Pb collisions at the LHC at an impact parameter $b=15$ fm. {\it Lower Panel:} Transverse momentum distribution of para-dimuonium produced in Pb + Pb collisions at the LHC at an impact parameter $b=15$ fm. The dashed lines are calculations following the equivalent photon method, as described in Ref. \cite{bn02}.}
\label{dpd}
\end{center}
\end{figure}

The impact parameter dependence of the PM production mechanism depends on the function $K_0$ which decays exponentially at large  impact parameters. The largest production probability occurs at the smallest possible impact parameter, $b\sim 2R_{\rm Pb}$, where $R_{\rm Pb}$ is the matter radius of Pb. The argument of $K_0$ becomes at this impact parameter approximately equal to $\xi = 4m_\mu R_{\rm Pb}/\gamma \sim 1/100 \ll 1$, so that the production probability  at small impact parameters scales as $\ln^2 \xi$. Thus, the production probability decays rather slowly, like $\ln b$, up to an impact parameter of the order of $b\sim 100$ fm, where an abrupt exponential decay of the form $\exp(-\xi)$ sets in.  This means that the compact PM will be produced when the ions collide with an impact parameter in the range $10 \ {\rm fm} \lesssim b \lesssim 100$ fm.

We use the relationship $y=\ln [(P_0+P_z)/2m_\mu]$ and $dP_z = 2m_\mu e^y dy$ to express Eq. \ref{dPdm} in terms of the rapidity $y$.  
In Fig \ref{dpdptdy} we show the momentum-rapidity distribution $d{\cal P}_{dm}/ dP_T dy$ (scaled to unity at $P_T=0, \ y=0$)  for PM production  in a Pb + Pb collision at the LHC at an impact parameter $b=15$ fm, which is larger than the sum of the radii of the colliding ions.  The momentum-rapidity distribution is strongly peaked at $P_T=0$ and is spread over rapidities in the range $|y| \lesssim 5$.

In Fig. \ref{dpd}, upper panel, we plot the rapidity distribution of PM produced in Pb + Pb collisions at the LHC at an impact parameter $b=15$ fm. The solid line corresponds the calculation using Eqs. (\ref{dPdm}-\ref{fob2}), whereas the dashed line is obtained using the equivalent photon method, Low's formula, and the corresponding $\gamma\gamma$-width for the dimuonium decay, as explained in Ref. \cite{bn02}. In the lower panel of the same figure we plot the transverse momentum distribution for the same reaction. 
The momentum distribution drops exponentially with $P_T$ (note the logarithm scale) peaking at $P_T=0$, as expected, and dropping by 8 orders of magnitude from $P_T=0$ to $P_T = 5$ fm$^{-1}$ (1 GeV/c). These values are within the range of possible experimental measurements with detectors such as the ATLAS and the CMS detectors existing at the LHC.  

The computations employing the equivalent photon approximation (EPA) yield slightly higher values, reaching up to $15\%$ more, compared to those obtained using Eqs. (\ref{dPdm}-\ref{fob2}). These discrepancies are most pronounced at small $P_T$ values, with EPA results consistently exceeding those from Eqs. (\ref{dPdm}-\ref{fob2}) across nearly all $y$ values. This comparative analysis highlights that EPA exhibits limitations in accuracy within certain regions of the phase space. This outcome aligns with expectations and has been comprehensively investigated in Ref. \cite{bn02}, which demonstrates that EPA accuracy is contingent on excluding specific regions of the phase space, such as instances involving small longitudinal and transverse momentum transfers distributed throughout the integrand \cite{bn02}.

From Fig.~\ref{dpd} we see that a significant fraction of the particles may have 
$ y \simeq 3$. Assuming that they have a transverse momentum much smaller
than their mass, $P_T << m$, their approximate energy will be            
$E = \sqrt{P_T^2 + m^2} \, cosh \, y$. Substituting the numerical values     
we find $E \simeq 2$ GeV in the CM frame. The two photons resulting from the  
dimuonium decay would have 1  GeV each and move in the beam direction.  
Forward photons with 
energies in the GeV region have been measured by the LHCf collaboration
\cite{lhcf}. Measuring
diphotons, they were able to reconstruct the parent $\pi^0$s and determine
their momentum distribution. Since the dimuonium and the $\pi^0$ have a similar
mass and decay into two photons, one could try to look for the dimuonium with
the same techniques. The important changes would be: 
i) the LHCf detector would have to be used in UPCs and 
ii) it should be adapted to cover smaller rapidities,
e.g. $2 < y < 4$, which is the rapidity range covered by the LHCb collaboration. 
It is important to observe that, based on the findings depicted in Fig.~\ref{dpd}, dimuonium production is probable at rest, indicating $y=0$ and $p_T=0$. Implicit in this observation is the assumption that, if the bound state decays isotropically, the resulting decay photons do not necessarily have to be detected solely in the forward direction.

To obtain the differential cross sections, we integrate Eq. \ref{dPdm} over impact parameters as
\beq
{d^2\sigma \over dP_z dP_T} = 2\pi \int b db \,  {d{\cal P}_{dm}(P_T,P_z,b)\over dP_T dP_z} S^2_{abs} (b), \label{d2sdpz}
\eeq
where $|S_{abs}|^2$ is the survival probability written as the square of the scattering matrix, introduced here to enforce absorption at small impact parameters \cite{BN93}, and
\beq
S_{abs} (b) = \exp\left[ -{\sigma_{NN} \over 4\pi}  \int_0^\infty dq q \, \tilde{\rho}_A^2(q) J_0(qb) \right], \label{smatrix}
\eeq
where  $\sigma_{NN}$ is the nucleon-nucleon cross section,  $\tilde{\rho}_A(q)$ is  the Fourier transform of the densities of the colliding ions and $J_0$ is the cylindrical Bessel function of zeroth order. 
For Pb we use a Woods-Saxon density with radius parameter 6.63 fm, and diffuseness equal to 0.549 fm. At the LHC energies the nucleon-nucleon cross section is  $\sigma_{NN} = 90 $ mb \cite{PDG}.  The S-matrix in Eq. \ref{smatrix} increases from 0 to unity very fast around two times the radius parameter. The rapidity and $P_T$ dependence of the cross sections display a similar pattern as the probability distributions  presented in Figs. \ref{dpdptdy} and \ref{dpd}. The total integrated cross section for Pb + Pb collisions at the LHC is found to be 2.32 $\mu$b. This cross section is for PM production in the ground state. The total cross section for production in any PM excited state with principal quantum number $n$ is increased by a factor $\sum_n 1/n^3= 1.202$, i.e. by about 20\%, reaching $\sigma_{PM} = 2.86 \ \mu$b at the LHC.  The knowledge of the total cross section gives us a possibility to estimate the PM production rate. The PM production event rate at an LHC nominal luminosity of $10^{34}$ particles/cm$^2$/s is therefore about $10^{28} \times 2.86  \times 10^{-6} \times 10^{-24} = 28.6\times 10^{-3}$/second  per high luminosity experiment, and only about $0.906 \times 10^{-9}$ events per beam crossing, according to the beam properties at the LHC \cite{Rho23}. One could benefit from the much larger luminosity for p-p collisions in the collider, larger by a factor $10^6$ as compared to Pb-Pb collisions. But the $Z^4$ enhancement due to ion charge (see Eq. \ref{dPdm}) more than compensates for the smaller luminosity of the Pb beams. Therefore, Pb + Pb collisions are the best reactions to produce the dimuonium at the LHC based on the number of events available.

\section{Discussion}

The observation of the dimuonium can be achieved by  searching for X-rays from  Bohr transitions such as $2P \rightarrow 1S$ (probably too difficult to measure in the LHC environment due to its small transition energy), their decay lengths, 2P lifetimes,  and other possible transition energies. The PM decay length after being created is about  $\lambda = c\tau_{PM} = 0.181$ mm. It decays by two photons each with energy $E_\gamma \sim m_\mu \sim 100$ MeV. For a photon with this energy, pair production will be the dominant photon interaction in most materials.  A reconstruction of the tracks of the charged $e^+$ ($e^-$) as they pass through a series of trackers, the $\gamma$-ray direction and therefore its origin where the dimuonium is produced can be inferred. Given the current limitations of LHC detectors, it is presumed that none of these experiments can be conducted using existing technology. However, a straightforward observation of dimuonium could serve as a pivotal step, opening up the potential for its comprehensive study through alternative experimental methods. This parallels the situation following the initial production and detection of antihydrogen, which paved the way for further investigation and insights \cite{Baur1996}.  

A small fraction of the dimuonia  will have large velocities and if they interact with matter, they will be dissociated  due their  Coulomb interaction with the nuclei within a thin material layer of the material with a cross section of the order of $Z^2 \times 10^{-23}$ cm$^2=10\times Z^2$ barns \cite{BB89}.  This is a large cross section, implying a dissociation length for a substance with number density $n$ given by  $\lambda \sim 1/n \sigma \sim 1$ mm, for a typical material with $Z \sim 30$. This value is only about a factor 10 larger than  the dimuonium decay length. Therefore, the observation of the dimuonium will also need to consider the detection of free muons. This poses a problem as the production cross section of free $\mu^+\mu^-$ pairs are orders of magnitude larger \cite{BB87} than that for the production of the dimuonium in heavy ion collisions. Thus, most free muons will be produced directly and will pose a large background for the observation of the dimuonium dissociation process. The interaction of the dimuonium with matter would be of great interest for QED. It has been predicted \cite{Nem81,LP81} that the probability of observing a positronium in a bound state after its passage through a thin layer of matter is inversely proportional to the layer thickness $L$. This positronium ``transparency'' takes place, when $L$ is much smaller than characteristic internal positronium time Lorentz dilated in laboratory system. This phenomenon deviates from the typical exponential law for the survival probabilities of charged particles in matter. The physical reason for this effect is that the positronium  fluctuates in different excited states during its passage through the slab of matter \cite{Zak87}. The observation of a relativistic dimuonium interacting with matter would further help understanding this QED property with implications to quantum field theory in general.

Assuming that one tries to identify experimentally the PM via its decay into two gamma rays, it is important to check the $\gamma$-background which can arise, e.g., by Bremsstrahlung, or by light-by-light scattering. Production of Bremsstrahlung photons at the LHC was estimated in Ref. \cite{BB89}. We consider the emission of photons with energy $E_\gamma = m_\mu$ within a window $\Delta E_\gamma \sim 10^{-2} m_\mu$ corresponding to a 0.01 accuracy in the measurement of a $\gamma$s from the dimuonium decay. The calculated Bremsstrahlung cross section if found to be smaller than the cross section for the PM production by a factor $10^{-4}$. This is because the ion mass, $m_{\rm Pb}$, is too large to be ``shaken'' appreciably as it passes by another ion moving in the opposite direction at impact parameters $b>2R_{Pb}$. 

Light-by-light scattering during a heavy ion collision, was also estimated in Ref. \cite{BB88} using the Delbr\"uck elastic scattering of photons by photons as input. In this model, the cross section for light-by-light scattering events leading to high energy gammas $E_\gamma \gtrsim m_\mu$, is given to leading order by integrating the Delbr\"uck cross section folded with equivalent photons around a window with uncertainty $\Delta E_\gamma = m_\mu/100$. One obtains
\begin{equation}
    \sigma_D \sim 2.54 \times 10^{-2} Z^4\alpha^4 r_\mu^2  \ln^3 \left( { \gamma \over m_\mu R_{\rm Pb}}\right) ,
\end{equation}
where  $r_\mu=e^2/m_\mu$ fm is the classical muon radius. For Pb + Pb collisions at the LHC, the result is $\sigma_D\sim 30$ nb, that is, two orders of magnitude smaller than the PM production cross section. Therefore, the direct photon production mechanisms such as  Bremsstrahlung or light-by-light scattering will not lead to substantial backgrounds in the detection of the dimuonium decay.

Another potential source of $\gamma$-ray background stems from the production of $\pi^0$ particles. The direct production of a single $\pi^0$ is not a significant concern, as it manifests as a distinct narrow peak at an energy corresponding to half the mass of a pion. However, the production of two pions can result in a broader spectrum of $\gamma$ energies that warrants careful consideration. This particular production mechanism has been examined in detail in Ref. \cite{KWS13}, where the calculated cross sections for this process were found to be on the order of a few tens of nanobars within the region of interest -- up to 1 GeV gammas. It is important to note that these cross sections are considerably smaller than our predictions for $\gamma$ production through the process of dimuonium production and decay in ultraperipheral collisions (UPCs).

The proposal we make in this work is to use the LHC facility at CERN to produce the dimuonium for the first time. The facility already exists, is well equiped, and has experience wih UPCs to produce particles in one- and two-photon exchange processes. We have shown that the production of the dimuonium at the LHC could be within experimental reach.  But one also needs to be realistic presently at the LHC one might not be able to resolve  low-energy transition such as the 2P-1S transition. Many other processes compete simultaneously with the UPC, with events induced by slow neutrons, cosmic rays and radiation from decaying materials. One has to rely on innovative ideas, similar to the ones developed in the past two decades to detect events generated by UPCs.

Perhaps the   most basic and important question to consider once leptonic atoms are studied, is how one accurately calculates a bound-state in QED starting from Feynman tree diagrams. This question is not as  easy to answer as that of QED treatment of free particles, because it relates to a bound system in which Coulomb exchange is not a small effect. To calculate the formation of a bound state a special treatment like a non-relativistic quantum electrodynamics (NRQED) approach is often used. Many studies of the positronium have contributed to the development of bound-state NRQED \cite{Kars02,Kars05}. The production and decay of the dimuonium will certainly be an important part of this study.

\section{Conclusions} 

To conclude, we would like to emphasize that our calculation is
very clean. It is a step along the direction of high precision QED  and 
tests of beyond the standard model physics. Our goal was to provide a 
theoretical guidance that can be used by experimentalists either
now or in the future using the magnitude of cross sections and the phase space
exploration in $y$ and $P_T$ so that they can judge if their complex 
detector environment is adequate to assess the predicted information.

\bmhead{Acknowledgments}
The authors acknowledge support by the U.S. DOE grant DE- FG02-08ER41533 and the Helmholtz Research Academy Hesse for FAIR. This work was  also partially financed by the Brazilian funding agencies CNPq, FAPESP and by the INCT-FNA and by the IANN-QCD network.

\hspace{1.0cm}

\end{document}